%% file: CSCW25_Sastha_TIES_CR.tex
\documentclass[acmsmall]{acmart}

\usepackage{makecell}
\usepackage{tabu, booktabs}
\usepackage[utf8x]{inputenc}
\usepackage{lmodern,textcomp}
\usepackage{tfrupee}
\usepackage{lmodern}
\usepackage{longtable}
\usepackage{adjustbox}
\AtBeginDocument{%
  }

\setcopyright{acmlicensed}
\copyrightyear{2025}
\acmYear{2025}
\acmDOI{XXXXXXX.XXXXXXX}

\acmConference[CSCW '25]{The 28th ACM SIGCHI Conference on Computer-Supported Cooperative Work \& Social Computing}{October 17--23, 2025}{Bergen, Norway}
\acmISBN{978-1-4503-XXXX-X/18/06}

\received{October 2024}
\received[revised]{April 2024}
\received[accepted]{August 2025}

\begin{document}

\title{Leveraging Familiarity with Television to Enrich Older Adults' Engagement and Wellbeing: A Feasibility Study Using Video Probes}
\renewcommand{\shorttitle}{\textit{TV as an Interactive Engagement System} for Older Adults}

\author{Shyama Sastha Krishnamoorthy Srinivasan}
\affiliation{%
  \institution{IIIT-Delhi}
   \city{New Delhi}
  \country{India}}
\email{shyamas@iiitd.ac.in}

\author{Mohan Kumar}
\authornote{Both authors are advisors of this research work.}
\affiliation{%
  \institution{RIT}
   \city{Rochester, New York}
  \country{US}}
\email{mjkvcs@rit.edu}

\author{Pushpendra Singh}
\authornotemark[1]
\affiliation{%
  \institution{IIIT-Delhi}
   \city{New Delhi}
  \country{India}}
\email{psingh@iiitd.ac.in}

\renewcommand{\shortauthors}{SS K Srinivasan, M Kumar, P Singh}

\begin{abstract}
The shift away from multigenerational families to nuclear families in India has created a growing need to support older adults living independently. While technology can help address this gap, older adults' limited exposure to newer technology restricts the adoption of such solutions. However, they remain comfortable with long-standing technologies like television (TV). This study explores their daily technology usage and challenges, aiming to determine whether TV can be leveraged to improve their quality of life. We examined how TV systems could be enhanced to assist older adults with tasks such as staying connected, receiving health alerts, and ensuring security. Using a participatory design approach, we developed video probes using the prototype of the TV-based application and interviewed 27 older adults to assess its acceptance and usability. The findings validate our hypothesis, as the strong interest in a TV-based solution underscores a preference for familiar technology to support security, independence, and wellbeing.
\end{abstract}



\maketitle

\input{1_intro}
\input{2_related}
\input{3_method}
\input{4_findings}
\input{5_discussion}
\input{6_conclusion}


\begin{acks}
\end{acks}

\bibliographystyle{acm}
\bibliography{references}

\end{document}

%% file: 1_intro.tex
\section{Introduction}

The global increase in life expectancy has led to a notable rise in the older adult population, projected to grow significantly in the coming years \cite{he_goodkind_kowal_2015, gaigbe-togbe_bassarsky_gu_spoorenberg_zeifman_2022}. In India, the comparatively low retirement age \cite{retirement_age_increase_1998, retirement_age_2011} combined with the trend towards nuclear families has resulted in older adults spending more time independently, often without close familial support (Census of India \cite{giridhar_demographics_2014}). As post-retirement spans extend and family structures change, the need for support systems that promote positive aging \cite{gergenPositiveAgingNew2001}—by fostering well-being, independence, and active engagement in society—has become increasingly vital \cite{ugargol_care_2016, adlakha_neighbourhood_2020}. Some initiatives have been pursued to address this need through elder-friendly neighborhood designs and specialized housing communities, predominantly in urban areas \cite{adlakha_neighbourhood_2020, phdcciadmin_senior_2021}. However, these solutions often entail high costs and limited accessibility, potentially reducing their applicability across broader older adult populations.

The advent of technology presents new opportunities to augment these support systems, providing practical means to improve the everyday lives of older adults \cite{10.1145/3411764.3445702, 10.1145/3170427.3170641, 10.1145/3274372}. Effective technological interventions require a nuanced understanding of older adults' unique needs and preferences, which differ across diverse demographics as shown by prior scholarship \cite{10.1145/3491101.3503745, 10.1145/3290607.3299025, 10.1145/3479506}. The challenges that older adults in India face towards digital use, are of two forms: (a) It is \textit{extrinsic} in the form of increased risk through scams, frauds, and other illicit activities targeting them \cite{sapkaleHowSeniorCitizens, chattopadhyayCYBERCRIMESELDERLYPEOPLE2024}; (b) It is \textit{intrinsic} in the form of knowledge gaps, cognitive loads, and an inherent nature to seek simplistic lifestyles.  While smartphone adoption among Indian seniors is increasing, many face challenges adapting to technology they did not grow up with, potentially due to limited learning resources, financial constraints, or health conditions affecting dexterity and mobility \cite{Nair2022TheAA, geriatrics7020028}. Even with smartphones, their applications remain limited in diversity and features, as observed by Sharifi et al. \cite{10.1145/3604269}. In contrast, television (TV), a familiar and established medium in Indian households \cite{shitak_2011}, has evolved into smart TVs, creating a viable avenue for technological interventions designed to support older adults.

The objective of this research is to address two primary questions: (1) \textit{What essential support factors do older adults perceive as critical for their aging experience?} and (2) \textit{How can TV be leveraged to assist older adults in daily tasks and enrich their quality of life?} To ground these questions, we first conducted an initial formative study aimed at exploring the aging experience of older adults in urban India. The study consisted of a survey with 60 respondents (42 males, 16 females) and 24 follow-up unstructured interviews (17 males, 7 females) across Delhi, Chennai, Bangalore, and Kanpur (spanning the north and south of India). This study helped us refine \textit{RQ.1} by identifying key factors such as health, social engagement, security, independence, and support that are vital for enhancing quality of life. Furthermore, the formative work confirmed our original hypothesis that TV (specifically any TV capable of running an application on its own or with augments like Roku, Firestick, Apple TV, and more) represents a viable medium for technological intervention\textemdash given its widespread presence and accessibility in Indian households, even among those facing challenges with smartphones. Building on these insights, our main study adopted a participatory design approach to design, develop, and explore the acceptance of a TV-based intervention. This phase involved conducting participatory design sessions with a subset of older adults from the formative study to create and iteratively refine prototypes with varying fidelity to develop \textit{video probes}\textemdash demonstration of prototypes of a TV-run application. These sessions helped us identify the features and interaction modalities most beneficial to the target users. Following this, for the interaction and evaluation of \textit{RQ.2}, a new cohort of 27 participants (14 males, 13 females) from Chennai and Tiruchirappalli (with different living arrangements) were engaged in semi-structured preliminary interviews to understand their life experiences and technology familiarity. Participants then interacted with the video probes, followed by detailed interviews to gather feedback, thereby reaffirming and further elucidating the support factors highlighted in \textbf{RQ.1}. Throughout this paper, we will use TVs synonymous with TVs capable of running an application.

Our analysis revealed factors such as emphasizing health, social engagement, security, independence, and support as crucial for enhancing older adults' quality of life (expanded in the findings). Our contributions entail empirical insights into methods for leveraging TV as an interactive platform to support older adults, recommendations for fostering social engagement through technology, and evaluation of video prototypes' effectiveness in design and development for the older adult population.

%% file: 2_related.tex
\section{Related Work}

\subsection{Technology, Older Adults, and HCI}

Existing research underscores the role of technology in enhancing older adults' social engagement and health management \cite{charness_aging_2009, pilotto_technology_2018, wang_technology_2019, moore_older_2021}. However, adoption relies not only on perceived utility but also on simplicity and alignment with established routines, as older adults exhibit preferences for technologies that are both familiar and accessible \cite{czaja_factors_2006, lee_perspective_2015}. Studies reveal that adoption patterns vary widely based on interface usability, perceived value, and cost-effectiveness \cite{czaja_social_2021}.

Intervention studies have explored various technological mediums aimed at enhancing social connectedness for older adults. For instance, Czaja et al. \cite{czaja_social_2021} introduced the PRISM system, which accounted for critical factors in technology adoption through targeted trials addressing loneliness. Video calls as a medium to reduce isolation were explored by Zamir et al. \cite{zamir_video_calls_2018}, while Cun Li \cite{10.1145/3294109.3302957} and Daniel et al. \cite{10.1145/3374920.3374922} analyzed intergenerational storytelling and emotional journaling. However, these interventions often remain highly specific to the contexts in which they are applied, particularly for mental and psychological wellbeing. They may not be transferable to populations with minimal exposure to advanced technology.

There has also been significant interest in voice assistant-based interventions, seen as potentially transformative for healthcare and quality of life \cite{10.1145/3441852.3471218, 10.1145/3597638.3614501, 10.1145/3597638.3608378}. Nonetheless, challenges remain, especially in adapting these interventions to diverse settings, such as India, where language diversity and socioeconomic disparities present unique obstacles. The work of Varella et al. \cite{technologies10010008} on adaptable intervention tools, such as online interfaces and helper bots, highlights these obstacles, particularly regarding infrastructure and support limitations.

Beyond these challenges, older adults' adoption barriers are influenced by a range of physical, cognitive, and socioeconomic factors \cite{olson_diffusion_2011, moxley_factors_2022, wilson_understanding_2023}. Declines in cognitive and motor skills, along with anxieties around complex interfaces, remain significant deterrents to engagement \cite{tams_modern_2014, lee_perspective_2015, mace_older_2022, wilson_understanding_2023}. Socioeconomic factors, including digital literacy, financial barriers, and limited access to technical support, further compound these challenges \cite{yusif_older_2016}.

Within the Indian context, these barriers are particularly prominent given generally lower levels of digital literacy and limited technology exposure \cite{yusif_older_2016}. Familiarity plays a crucial role in technology acceptance as seen in prior scholarship \cite{wade_bridging_2002, wu_bridging_2015, martins_van_jaarsveld_effects_2020}. With TV already integrated into daily life \cite{shitak_2011}, it may serve as an effective medium for intervention, integrating social and health-related activities without the complexities associated with newer technologies. Research has shown that TV has historically been overlooked as a medium for interactive interventions due to concerns about excessive viewing habits \cite{buchanan_reducing_2016}. However, it may offer potential in this demographic when applied with moderation to avoid cognitive decline associated with excessive use \cite{fancourt_television_2019}.

Emerging uses of smart TVs have introduced possibilities for social and health-oriented interventions among older adults, particularly for those less inclined to adopt mobile or web-based solutions \cite{Nair2022TheAA, geriatrics7020028}. The FoSIBLE project, for example, used TV as a platform for social engagement, incorporating functionalities like video calling and health monitoring \cite{alaoui_increasing_2012}. Davoodi et al. \cite{davoodi_interactive_2021} examined the potential of TV for reducing loneliness by offering activity-based episodes focused on wellbeing. However, these interventions remain largely untested in non-Western contexts such as India, where constraints on affordability and infrastructural support limit their applicability.

This study addresses this gap by evaluating a TV-based intervention system designed to enhance the quality of life for older adults in India, integrating social and health-related features into a familiar platform to facilitate ease of use and adoption.

\subsection{Designing and Developing for Older Adults}

Quality of life (QoL) for older adults is an essential multidimensional construct encompassing physical, psychological, and social aspects. The World Health Organization defines QoL as an individual's perception of their position in life, shaped by expectations, goals, and cultural context in their \textit{World Health Organization Quality of Life} (WHOQoL) Questionnaire \cite{whoqol_old}. Studies by Rowe and Khan \cite{rowe1997successful} and investigations in India by Samanta et al. \cite{samanta_living_2015} and Srivastava et al. \cite{srivastava_pursuit_2021} indicate that social, familial, and financial stability are central to wellbeing in older adults. These necessitating designs foster independence as they live alone or in smaller family units \cite{ausubel_older_nodate}.

A participatory design approach has proven crucial in developing technologies that align with the needs and preferences of older adults \cite{10.1145/2207676.2208570, 10.1145/3536169.3537791}. This methodology not only ensures that technologies remain accessible but also involves older adults in co-creation (as can be observed from prior work \cite{10.1145/2513383.2513451, 10.1145/3033701.3033717, 10.1145/3613904.3641887}), contributing to the usability and acceptability of technologies, such as voice assistants and smart home devices \cite{10.1145/2466627.2466652, 10.1145/3613904.3642595, 10.1145/3643834.3661536}.

This study extends the existing literature by utilizing participatory design to develop a TV-based system for urban Indian older adults. Through co-design workshops, participant feedback on system prototypes informed iterative design improvements, enabling a refined, user-centered intervention. Video probes are valuable for engaging users in designing and evaluating new technologies, mainly when traditional usability testing is not feasible \cite{10.1145/191666.191712}. In this work, video probes were utilized to capture user engagement in a natural setting, with prior literature such as the work from Mackay and Fayard \cite{10.1145/632716.632790} and Pang et al. \cite{10.1145/3411764.3445702} demonstrating the credibility for obtaining qualitative insights on technology use preferences in real-life contexts. This technique provided valuable data on user interaction without requiring real-time interface adaptation, allowing for the collection of nuanced feedback on this TV-based intervention for older adults in urban India.

%% file: 3_method.tex
\begin{figure}[ht]
    \centering
     \includegraphics[width=.96\linewidth]{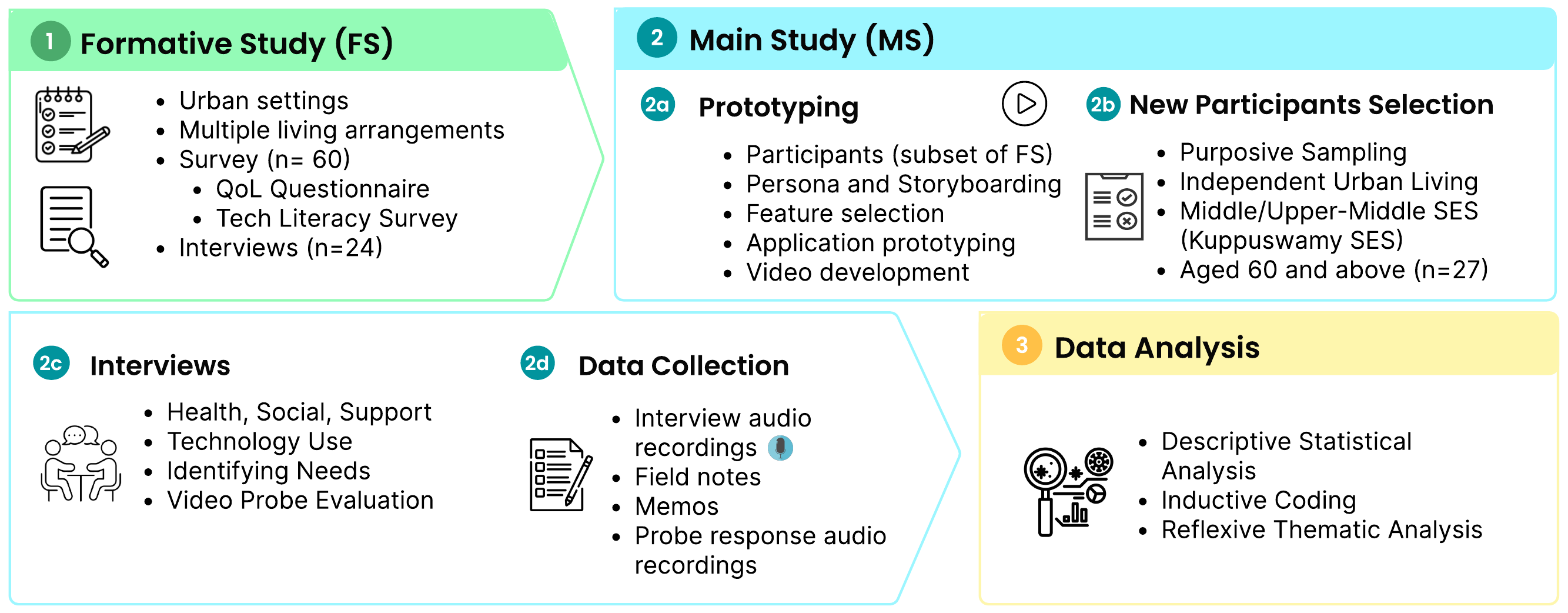}
     \Description[The detailed workflow process of the study]{The workflow process diagram in Figure 1 shows the research flow beginning from the formative study and literature survey, with findings indicating the factors older adults consider essential for their aging experience and TV as a viable medium for support. This leads to the prototyping part of the main study including personas and scenarios for the video probe development through a participatory design-based approach. It is followed by the main study design, including participant selection, interviews, and data collection. Finally, the methods used for data analysis are listed.}
     \caption{\label{Figure 1} Workflow breakdown of the research study.}
\end{figure}

\section{Methodology}

Our research explores the motivations and experiences of independent older adults in urban India, focusing on digital technology and TV as intervention tools. Using a sequential, mixed-methods approach, we conducted a formative study (with a survey and unstructured interviews), followed by a main study (with participatory design workshops, interviews, and video probe evaluations), each stage building on the previous one (Fig. \ref{Figure 1}). The formative study validated our hypothesis of using TV as the intervention medium, leading to the main study, where rapid prototyping helped refine key features for enhancing security, social connectedness, health, and wellbeing. Video probes were then used with new participants to assess both their lifestyles and the feasibility of a TV-based application. Detailed methodology, including participant demographics, recruitment, data collection, and analysis, follows. While this is the methodology of our research, brief descriptions of outcomes from the formative study are also presented, as it helps readers understand the choice of design elements for the main study.

\begin{table}[ht]
\centering
\begin{adjustbox}{width=\textwidth}
\begin{tabular}{l  c  c c  c c  c c  c c c}
\toprule
\textbf{Method} & \textbf{Total} &\multicolumn{2}{c}{\textbf{Gender}} & \multicolumn{2}{c}{\textbf{Age}} & \multicolumn{2}{c}{\textbf{Family Size}} & \multicolumn{3}{c}{\textbf{Education}}\\
\cmidrule{3-11}
& & \textbf{Male} & \textbf{Female} & \textbf{Range} & \textbf{Mean} & \textbf{1-2} & \textbf{3+} & \textbf{High} & \textbf{College} & \textbf{Advanced/}\\
& & & & & & & & \textbf{School} & \textbf{Graduate} & \textbf{Doctorate}\\
\midrule
\textbf{Survey} & 60 & 43 & 17 & 60-88 & 66 & 14 & 46 & 13 & 18 & 29\\
\textbf{Open-ended} & 17 & 13 & 4 & 61-88 & 67 & 3 & 14 & 3 & 6 & 8\\
\textbf{survey questions} & & & & & & & & & & \\
\textbf{Interviews} & 24 & 17 & 7 & 61-88 & 68 & 6 & 18 & 5 & 8 & 11\\
\bottomrule
\end{tabular}
\end{adjustbox}
\caption{\label{Table 1}Demographics of the participants from various formative study components. \textit{*High school means the maximum schooling possible during their time ($11th$/$12th$ grade.); College degree means a bachelors diploma/degree; Anything more than 16 years of education is considered Advanced (post-graduate degree/diploma)).}}
\Description[The table listing participant statistics from various study components]{Table 1 lists the ranges and mean of age, gender, family size, and educational background for the three components of the formative study.}
\end{table}

\begin{table}[ht]
    \centering
    \begin{adjustbox}{width=\textwidth}
    \begin{tabular}{ l l l l l l l l l }
    \toprule
        \textbf{ID} & \textbf{Age} & \textbf{Family} & \textbf{Gender} & \textbf{Education} & \textbf{Capability} & \textbf{Attitude} & \textbf{Willingness} \\ 
        & & \textbf{Size} & & & & & \\\midrule
        FP1 & 61 & 1-2 & male & College Degree & SP and Laptop & Neutral & Medium \\ 
        FP2 & 88 & 3-6 & female & Advanced Degree & KP and TV & Negative & Low \\ 
        FP3 & 65 & 3-6 & female & High School & SP and Laptop & Positive & High \\ 
        FP4 & 75 & 3-6 & male & High School & SP and Laptop & Positive & Medium \\ 
        FP5 & 65 & 3-6 & male & College Degree & SP and Laptop & Neutral & Medium \\ 
        FP6 & 63 & 3-6 & male & Advanced Degree & KP and TV & Negative & Low \\ 
        FP7 & 61 & 3-6 & male & Advanced Degree & KP and TV & Negative & Low \\ 
        FP8 & 61 & 3-6 & male & Advanced Degree & KP and laptop & Negative & Low \\ 
        FP9 & 61 & 3-6 & male & College Degree & KP and TV & Negative & Low \\ 
        FP10 & 61 & 3-6 & male & College Degree & KP and TV & Negative & Low \\ 
        FP11 & 87 & 3-6 & female & Lower Primary & KP and TV & Negative & Low \\ 
        FP12 & 62 & 3-6 & male & Advanced Degree & KP and TV & Negative & Low \\ 
        FP13 & 63 & 3-6 & male & Doctorate & Most Technology & Neutral & High \\ 
        FP14 & 64 & 1-2 & female & Advanced Degree & SP and Laptop & Neutral & Medium \\ 
        FP15 & 67 & 1-2 & male & College Degree & KP and Computer & Positive & Medium \\ 
        FP16 & 66 & 1-2 & male & College Degree & Laptop, SP, Ipad & Neutral & High \\ 
        FP17 & 72 & 7+ & female & Lower Primary & SP and TV & Positive & High \\ 
        FP18 & 62 & 1-2 & female & College Degree & SP & Positive & Medium \\ 
        FP19 & 71 & 1-2 & male & Advanced Degree & KP & Negative & Low \\ 
        FP20 & 72 & 7+ & male & Advanced Degree & SP & Positive & Medium \\ 
        FP21 & 68 & 7+ & male & Advanced Degree & SP and TV & Negative & Low \\ 
        FP22 & 67 & 3-6 & male & Advanced Degree & SP & Neutral & High \\ 
        FP23 & 72 & 7+ & female & No formal Education & KP and TV & Positive & Medium \\ 
        FP24 & 77 & 7+ & male & Diploma & SP and TV & Negative & Low \\  \bottomrule
    \end{tabular}
    \end{adjustbox}
    \caption{\label{Table 2} Participant Demographics \& self-reported attitude and willingness towards technology use from the formative study. \textit{**legend: SP - Smartphone, KP: Keypad Phone; *High school means the maximum schooling possible during their time ($11th$/$12th$ grade.); College degree means a bachelors diploma/degree; Anything more than 16 years of education is considered Advanced (post-graduate degree/diploma)).}}
    \Description[The table listing participant demographics]{Table 2 lists details of age, gender, family size, educational background, attitude, and willingness to use technology for each participant by the uniquely assigned participant ID from the formative study.}
\end{table}

\begin{table}[ht]
\centering
\begin{adjustbox}{width=\textwidth}
\begin{tabular}{l  l  l}
\toprule
\textbf{Themes} & \textbf{Sub-Themes} & \textbf{Main Codes}\\
\midrule
\textbf{Factors Influencing} & Personal Well-being & Physiological Challenges\\
\textbf{Aging Experience} & & Psychological Challenges\\
& & Social Interactions\\
& & Family Support\\
\rule{0pt}{3ex}
& Living Arrangements & Facilities\\
& & Accessibility\\
\rule{0pt}{3ex}
& Outlook on life & Perspective on Aging\\
& & Family well-being\\
\midrule
\textbf{Technology Use, }& Attitude towards & Awareness\\
\textbf{Perception, }& Technology Usage & Interest/Disinterest\\
\textbf{and Challenges }& & Importance of Technology Use\\
\rule{0pt}{3ex}
& Willingness/Challenges & Ease of understanding\\
& to use Technology & Usability\\
& & Learning Curve\\
& & Reliability\\
& & Affordability\\
\rule{0pt}{3ex}
& Actual Adoption & Time and Cost\\
& & Lack of disruption in daily life\\
& & Convenience\\
\bottomrule
\end{tabular}
\end{adjustbox}
\caption{\label{Table 3}Coding chart for formative study (Using qualitative data from the survey's open-ended questions and interview's field notes).}
\Description[Thematic Analysis table.]{Table 3 lists the main themes, sub-themes, and the main codes obtained from the reflexive thematic analysis of the open-ended survey questions and interview transcripts from the formative study.}
\end{table}
\vspace{-8mm}

\subsection{Formative Study}
\begin{figure}[ht]
    \centering
     \includegraphics[width=0.48\linewidth]{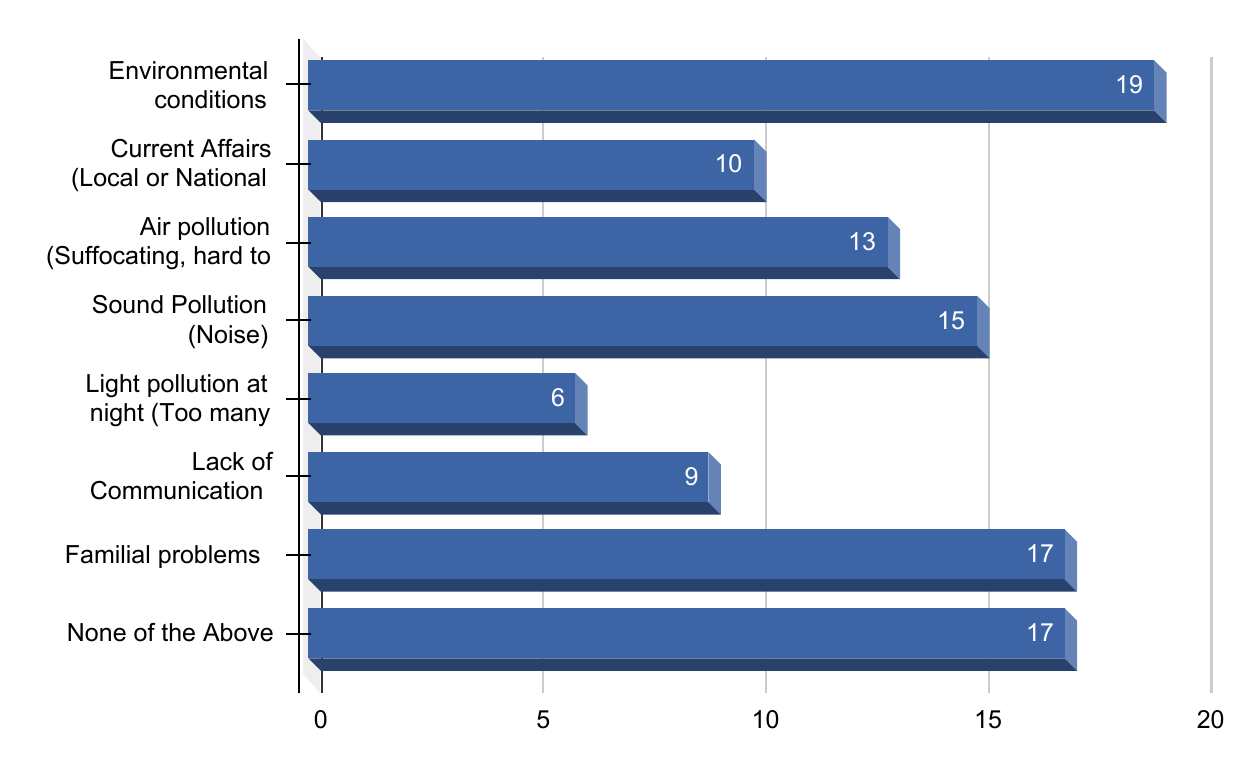}\hfill\includegraphics[width=0.48\linewidth]{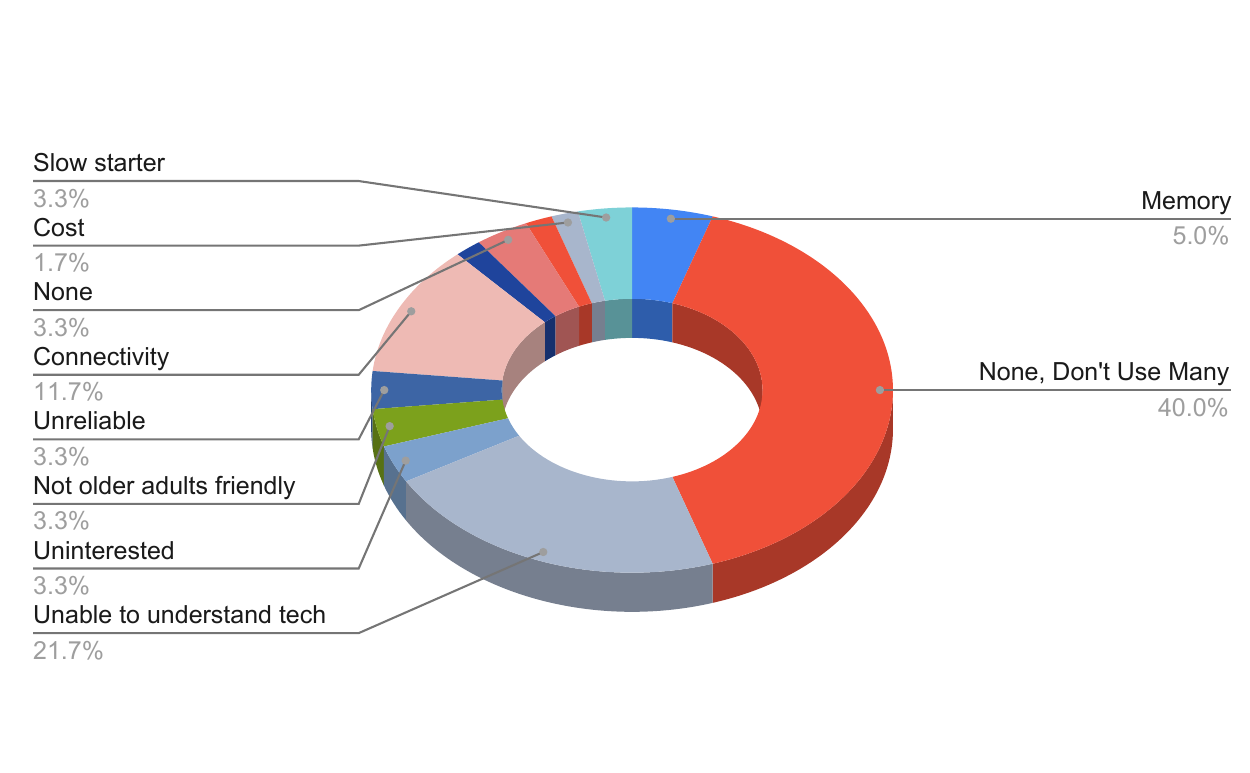}\hfill
     \Description[Bothersome factors and Challenges with Tech Use]{A pie chart made from the most common challenges faced by older adults in tech use, including the number of people who don't use many of the technologies listed in the survey.}
     \caption{\label{Figure 2} As listed by the participants in the survey: (a) Most bothersome factors observed by older adults, (b) The most common challenges in using technology.}
\end{figure}

The formative study investigated older adults' perceptions of aging, technology use, and barriers encountered in urban India. Our primary research questions were: \textit{"What factors do older adults in urban India consider important for a fulfilling aging experience?"} and \textit{"Can technology play a role in enhancing this experience?"} Conducted across four cities to cover the north and south of India (Delhi, Kanpur, Chennai, Bangalore), it comprised a survey (n=60) and 24 interviews (18 from Chennai and Bangalore, six from Kanpur and Delhi), with participants from multiple living arrangements (such as independent households, multi-family households, and community living) detailed in Table \ref{Table 1}. The survey contained two sets of questions. The initial questions were inspired by WHOQOL \cite{whoqol_old}, with questions like \textit{"Which of the following basic activities are you able to perform comfortably daily?" and "How would you rate your level of independence and control over your life?"} and more. The second set was focused more on the technological awareness, usage, and experience (in the form of a Likert scale) of digital technologies, from microwaves and washing machines to smart speakers and smart home systems. Participants aged 60 and above and from middle or upper-middle SES (according to the updated Kuppuswamy scale \cite{sood_modified_2022}) were recruited via snowball sampling from social platforms. A subset of the participants who filled out the survey were recruited for the interviews, where we explored further to understand some of their responses more qualitatively through unstructured interviews. The interview questions were along the lines of \textit{"Where do you live and who do you live with?", "How does your typical day look like?", "What technologies do you use and what are your feelings about them?"}, and more, depending on the flow of the interviews. Quantitative data (analyzed through descriptive statistics) highlighted challenges (Figure \ref{Figure 2}) with technological adoption due to substantial learning curves, limited reliability, and complex interfaces. The qualitative data were translated and transcribed into English, followed by collective examinations from the authors. Reflective thematic analysis \cite{clarke_successful_2013} on qualitative data about factors they considered essential for their aging experience underscored themes related to \textbf{connectivity, security, and health.} These findings established a foundation for creating a user-centered interface for our TV-based intervention, emphasizing ease of use and essential functionality. Table \ref{Table 2} provides demographic data, including participants' technology capability, general attitudes, and willingness to use technology. Table \ref{Table 3} presents the themes, sub-themes, and codes formed from the open-ended survey questions and the interview transcripts. Findings were drawn towards older adults' life experiences, which highlighted a generally low technology adoption, albeit with frequent TV use (22/24 participants) and a preference for straightforward, dependable interfaces, suggesting TV as a promising medium for intervention. The choice features of the TV-based application were considered, depending on the challenges listed by the participants, such as difficulty keeping in touch, declining health, and worries about security. These challenges were further analyzed while making storyboards and following through with multiple participatory rounds (one using figma prototypes) before developing the video probes for the main study.

\subsection{Leveraging \textit{TV as an Interactive Engagement System (TIES)}}
The primary goal of this intervention was to build support for the factors that older adults considered essential for their aging experience\textemdash staying in touch, security, and personal wellbeing. Insights from the formative study indicated that older adults in India are generally comfortable with TV due to familiarity and accessibility, aligning with prior studies on technology use among older adults \cite{peek_older_2016, anderson_tech_2017} and the holistic approach of HCI in aging \cite{10.1145/3290607.3299025}. Despite the potential of voice technology indicated by prior research and practice in the Western context \cite{10.1145/3359316, 10.1145/3373759}, the diversity of Indian languages and the ongoing development of regional language processing still at a nascent stage led us to focus on TV as a more familiar medium. Our intervention (\textit{TIES}) is a TV-based application that offers features like text/voice communication, reminders, video doorbell integration, video calls, and activity detection based on TV usage (or the lack of it), catering to the daily needs of older adults through interactive engagement solely through the TV. While we initially planned to use interviews to gather more feedback, one of the participants from the formative study suggested that they are from a generation of visual learners and showcasing these features in \textit{mock} demonstrations (maybe using videos) would help them grasp the workings better. This led us to consider the development of video probes that represent the features discussed in storyboards, showcasing how they operate under various pre-determined constraints. Through an iterative process (of $3$ sessions), we used participatory design to create prototypes, further developing into video probes that demonstrated the application's usability in various scenarios while prioritizing privacy and high-standard APIs during the design process. These probes helped gather participants' perceptions and feedback for further refinement in preparation for a pilot deployment. The prototyping details are further elaborated in the video probe design subsection.

\subsection{Video Probe Design}
Video probes were utilized to demonstrate the transformation of TV through \textit{TIES} in enhancing social engagement and wellbeing, capitalizing on the affordability of smart TV technology since Roku's \cite{news2008NetflixUnveils2008} 2008 debut. Utilizing Android OS, we created figma prototypes using personas and storyboards developed during participatory design sessions. Each persona represented distinct combinations of technological savviness and family proximity to reflect common user types. Based on participant feedback, we refined features and ultimately discarded a less-desired image gallery feature as participants found that to be trivial and a need-based feature. An example persona and storyboard are given below, which helped refine and develop feature prototypes later.

\begin{quote}
    \textit{\textbf{Persona 1:} “Ravi” is a 70-year-old retiree living alone in Chennai. He uses a keypad phone and watches TV for 2-3 hours daily. Ravi values staying in touch with his grandchildren and feels concerned about his health. The \textit{TIES} intervention offers him video calling options and health reminders integrated with his TV.}\\
    \textit{\textbf{Storyboard 1:} The video probe depicted Ravi receiving a medication reminder from his grandson while watching TV. He acknowledged the value of the personal interaction and appreciated that the reminders appeared in the center of the screen, pausing the program, ensuring he did not miss either.}
\end{quote}

\begin{figure}[ht]
    \centering
     \includegraphics[width=.96\linewidth]{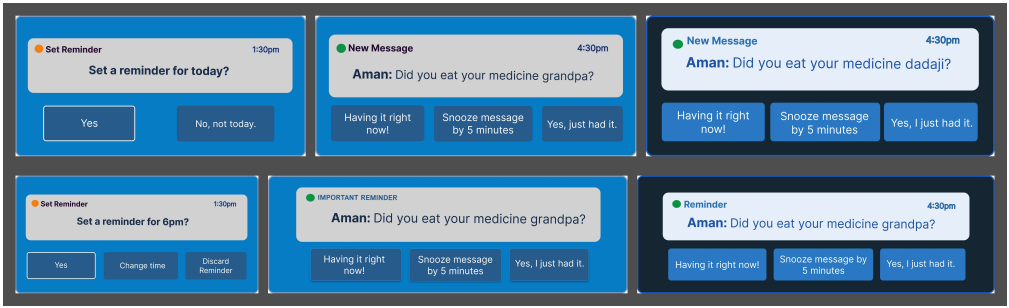}
     \caption{\label{Figure 3} Example of Various XML options (figma prototypes) shown to the participants for iterative feedback.}
     \Description[Exported Images from Figma]{The images in Figure 2 show examples of options shown to the participants during the participatory design phase regarding setting reminders and messaging, providing two alternative options in the top and bottom rows.}
\end{figure}

\begin{figure}[ht]
    \centering
     \includegraphics[width=.96\linewidth]{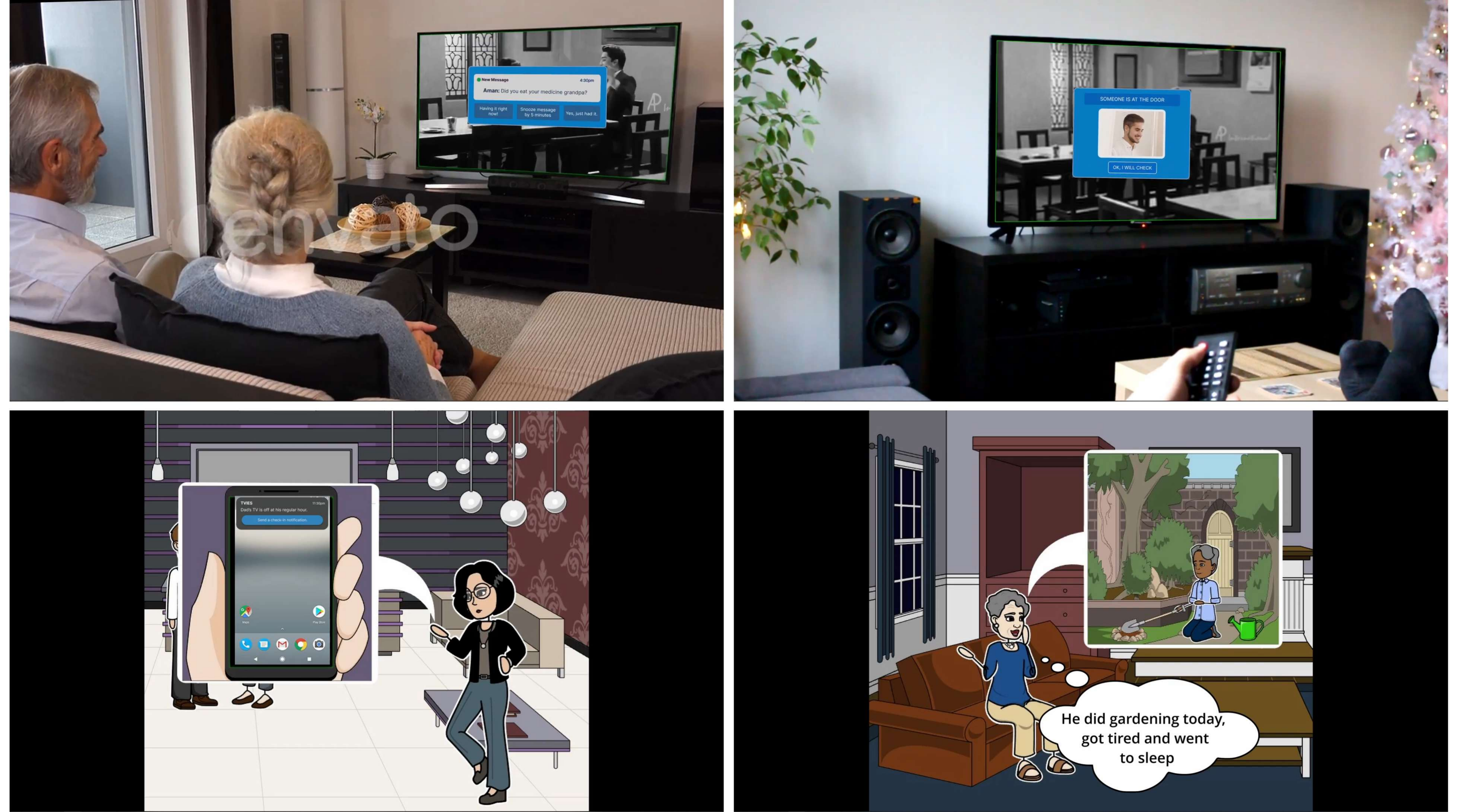}
     \caption{\label{Figure 4} Snippets from the developed video prototypes showcasing $4$ different features. The top left image showcases how a reminder will be displayed on the TV interface, followed by a video doorbell to its right. In the bottom row, two snippets showcase a scenario of the unusual absence of TV usage.}
     \Description[The snippets from the video prototype]{The images in Figure 3 show snapshots from the video prototype showcasing social connection and wellbeing monitoring features using the application prototype. The features included are reminders (from family/self-managed), video doorbell notifications, and activity detection through the unusual absence of TV usage.}
\end{figure}

\subsubsection{Participatory Design Iterations}
To get the design considerations for end users, $3$ participatory design sessions were conducted by recruiting the participants from the formative study who were willing to further provide their inputs. In the first session, we showed them the four distinct personas we had generated through the analysis of surveys and interviews. These personas relied on a combination of tech-savvy/non-tech-savvy and family living away/family living locally. The reasons for these combinations depended on the needs of individuals, depending on how quickly their family can be reached and their comfort level with operating technology. In the first round, using these personas, we sought feedback on what critical factors they believed could be addressed to improve their aging experience. Security, social connectedness, and sense of care were some of their primary concerns or goals. To that end, we created storyboards involving how TV can be utilized to augment their experience in these areas, and provided them to the participants in the second session, who provided further feedback on the way the interaction could happen using pen and paper diagrams on the positioning, font size, and appearance of the notifications. The final session was conducted post the figma prototype development.

\subsubsection{Prototypes and Video Probes Development}
Employing these selected use cases from the initial two participatory design sessions, we made the figma prototypes. Features included messaging, reminders, video calls, security, and privacy-conscious activity detection, recording only TV watch times. Also, we ensured the users had complete control over who could receive the data analytics by keeping everything within their local network during the development of the application. Once we designed the initial prototypes, we again showed them to the participants to gather feedback. First, the skeletal interactive XML (examples in figure \ref{Figure 3}) was shown to the participants during this process. Feedback from initial prototypes shaped the user interface and interaction modalities, culminating in a functional application demonstrated through storyboard videos. We ensured that feedback was obtained from the formative study participants to improve various aspects of these features, like text, the position of the notification, the duration of the interruption, and so on, before finalizing the probes for the main study. One of the recommendations arising from the participatory design sessions, which were incorporated during the bare functional backend application development process, was that all interactions with the TV application, including responding to notifications and recording voice messages, should be fully operational with a traditional smart TV remote buttons. After a few iterations of feedback incorporation, we developed the application's front end with a bare functional backend to create the storyboard videos (snapshots displayed in \ref{Figure 4}). These videos served as probes in the main study, with seven video probes provided as supplementary material. The selection of features was limited to those feasible in existing TV setups (aligning with formative study findings) requiring no additional cost, and the video calling and doorbell features were included after finding out it is possible to incorporate them within \rupee$5000$/\$$60$/€$55$.

\subsubsection{Video Probes Description}
All seven video probes were based on the finalized storyboards from the selected features. Each of the video probes consists of playing out a scenario when the TV is being used by older adults (or off at unusual times for one of the activity detection). While older adults are engaged in watching TV, the TV-based app sends prompts/notifications based on the feature, followed by relevant actions to be taken. The details of each probe are listed below.

\begin{itemize}
    \item Probe 1: Medication reminder from family on the TV screen
    \item Probe 2: Video Doorbell feed displayed for enhanced security
    \item Probe 3: Realtime messaging through TV interface (voice/text)
    \item Probe 4: Setting reminders and events with alarm feature
    \item Probe 5: Video calling with friends and family
    \item Probe 6: Unusual activity detection for unusual TV usage
    \item Probe 7: Unusual activity detection for unusual lack of TV usage
\end{itemize}

\begin{table}[ht]
    \centering
    \begin{adjustbox}{width=\textwidth}
    \begin{tabular}{ l l l l l l l l l }
    \toprule
        \textbf{ID} & \textbf{Age} & \textbf{Dyadic} & \textbf{Gender} & \textbf{Living} & \textbf{Education} & \textbf{Work} & \textbf{Perception} & \textbf{Use} \\ \midrule
        P1 & 68 & No & Male & Community & CA & Auditor & Positive & Negative \\
        P2 & 74 & Yes & Male & Community & High School & Banker & Positive & Negative \\
        P3 & 63 & Yes & Female & Community & High School & Homemaker & Positive & Negative \\
        P4 & 63 & Yes & Female & Individual & College Degree & GW & Positive & Negative \\
        P5 & 68 & Yes & Male & Individual & Advanced Degree & Teacher & Positive & Negative \\
        P6 & 74 & Yes & Male & Community & College Degree & Office Worker & Neutral & Negative \\
        P7 & 64 & Yes & Female & Community & High School & Homemaker & Neutral & Negative \\
        P8 & 68 & No & Female & Community & College Degree & GW & Positive & Positive \\
        P9 & 78 & No & Female & Community & College Degree & Teacher/Clerk & Positive & Positive \\
        P10 & 72 & Yes & Female & Community & High School & Homemaker & Neutral & Neutral \\
        P11 & 83 & Yes & Male & Community & College Degree & GW & Neutral & Neutral \\
        P12 & 83 & Yes & Female & Community & College Degree & Homemaker & Positive & Neutral \\
        P13 & 85 & Yes & Male & Community & College Degree & GW & Positive & Positive \\
        P14 & 76 & Yes & Female & Community & Doctorate & Professor & Positive & Positive \\
        P15 & 73 & Yes & Female & Individual & Doctor & Doctor & Positive & Positive \\
        P16 & 60 & No & Male & Individual & College Degree & GW & Positive & Positive \\
        P17 & 63 & No & Male & Individual & College Degree & GW & Positive & Neutral \\
        P18 & 81 & No & Female & Community & College Degree & Teacher & Positive & Neutral \\
        P19 & 70 & No & Male & Individual & College Degree & GW & Positive & Positive \\
        P20 & 80 & No & Male & Individual & College Degree & Banker & Positive & Neutral \\
        P21 & 61 & No & Male & Individual & High School & GW & Positive & Positive \\
        P22 & 64 & Yes & Female & Individual & College Degree & Teacher & Positive & Negative \\
        P23 & 70 & Yes & Male & Individual & College Degree & GW & Positive & Negative \\
        P24 & 73 & Yes & Male & Individual & High School & Entrepreneur & Positive & Neutral \\ 
        P25 & 67 & Yes & Female & Individual & College Degree & Teacher & Positive & Neutral \\
        P26 & 76 & Yes & Female & Individual & High School & Homemaker & Positive & Neutral \\
        P27 & 81 & Yes & Male & Individual & High School & GW & Positive & Positive \\ \bottomrule
    \end{tabular}
    \end{adjustbox}
    \caption{\label{Table 4} Participant Demographics \& self-reported orientation towards perception and use of TV intervention prototype from the main study. \textit{**legend: CA - Chartered Accountant, GW - Government Worker; *High school means the maximum schooling possible during their time ($11th$/$12th$ grade.); College degree means a bachelors diploma/degree; Anything more than 16 years of education is considered Advanced (post-graduate degree/diploma)).}}
    \Description[The table listing participant demographics]{Table 4 lists details of age, gender, type of living arrangements (individual/community), educational background, perception about the probe, and their willingness to use it for each participant by the uniquely assigned participant ID from the main study. It also has a column indicating whether or not the participant was part of dyadic interviews.}
\end{table}

\begin{table*}[ht!]
    \centering
    \begin{adjustbox}{width=\textwidth}
    \begin{tabular}{ l l l }
    \toprule
\textbf{Theme} & \textbf{Sub-Theme} & \textbf{Codes} \\
\midrule

\textbf{Changes in Daily Life} & 
Joy post-retirement & 
More time for family and spiritual \\
& & Enjoys relaxed life \\
& & Learning new things like languages \\
\rule{0pt}{3ex}

& Frustration with reduced mobility & 
concerned with reduced activity \\
& & Reluctant to travel due to health risks \\
\rule{0pt}{3ex}

& Adaptation to new routines & 
Community helps with daily chores \\
& & Adjusting to slower-paced living \\
\midrule

\textbf{Health and Wellbeing} & 
Routine health management & 
Takes medicine regularly\\
& & Uses a nebulizer twice daily \\
& & Has memory problems \\
\rule{0pt}{3ex}

& Coping strategies for & Regular check-ups help maintain health \\
& chronic conditions & Considers normal with age \\
\rule{0pt}{3ex}

& Emotional wellbeing sources & 
Daily meditation and prayer are essential \\
& & Spiritual content on TV provides peace \\
\midrule

\textbf{Community and} & 
Reliance on family and & 
Daughter helps with tech issues remotely \\
\textbf{Social Connectedness}& community support & Community assists with chores \\
\rule{0pt}{3ex}

& Preference for & 
Attends community functions if possible \\
& in-person interaction & Gated community provides security \\
\midrule

\textbf{Technology Use } & 
Some familiarity with & 
YouTube/OTT on TV for news/spiritual\\
\textbf{ and Resistance} & TV and smartphone & Enjoys using technology \\
& & Uses Whatsapp for social connectedness \\
\rule{0pt}{3ex}

& Resistance to new & 
Reluctant to use newer messaging apps \\
& technology and features & Wants control over notifs frequency \\
& & Thinks it is a distraction \\
& & Wants a simple life \\
\midrule

\textbf{Role of TV in } & 
Integration into spiritual & 
Uses TV for spiritual programs\\
\textbf{Routine and Autonomy} & and health routines & Finds health updates on TV useful \\
& & Thinks calling feature supports arthritis \\
\rule{0pt}{3ex}

& Preferences for & 
TV Health reminders must be minimal \\
& non-intrusive features & Dislikes unsolicited recommendations \\
\midrule

\textbf{Barriers and Facilitators} & 
Personal reluctance and & 
Prefers existing over new apps \\
\textbf{ to Adoption}& cognitive load concerns & Avoids complex features for peace \\
\rule{0pt}{3ex}

& Support from family and & 
Daughter sets up new features remotely \\
& community for adoption & Community assists with troubleshooting \\
\bottomrule

\end{tabular}
\end{adjustbox}
    \caption{\label{Table 5}Coding chart for main study (Using qualitative data from the interview's transcript \& field notes).}
    \Description[Thematic Analysis table.]{Table 5 lists the main themes, sub-themes, and the main codes obtained from the reflexive thematic analysis from the interview transcripts and the field notes taken during the interviews from the main study.}
\end{table*}

\subsection{Main Study}
The main study utilized video probes made with video prototypes to assess the impact of TV as an intervention medium on the wellbeing of independent residents of Tiruchirappalli and Chennai, India. Recruitment was done through purposive and snowball sampling, and participants were selected based on housing availability for older adults and the first author's knowledge of the area. Participants were chosen across varying living arrangements to capture diverse experiences, ensuring the sample included individuals both with and without close family networks nearby. This approach allowed us to explore varying levels of reliance on TV interventions based on proximity to caregivers. We ensured that different participants from those in the formative study were selected to avoid any unwanted bias.

Demographics such as age, education, occupation, perception towards the TV intervention, and willingness to use it were recorded (Table \ref{Table 4}), with some interviews conducted dyadically with spouses (A field in the table signifies whether a participant was part of a dyadic interview). Dyadic interviews offered valuable insights into how family dynamics affect technology use, particularly joint decision-making and adoption. Individual interviews, in contrast, provided a deeper understanding of participants' motivations and challenges. Each session lasted up to an hour, including interviews and a video probe conducted in a single sitting. The first author led the sessions, creating field notes and audio recordings to identify themes and ensure comprehensive data coverage, focusing on participants from various living arrangements and middle to upper-middle socioeconomic statuses.

Semi-structured interviews probed participants' lives pre- and post-retirement, examining changes in daily activities, social interactions, health, and technology use. These interviews paralleled the formative study's themes, ensuring the selection of appropriate demographics for assessing a TV-based intervention. Before showing the video probes, the participants were briefly explained what the probe was about, and during the probe, the participants were free to stop and ask questions if they had any. Participants viewed all seven video probes\textemdash excluding the one with a trivial picture gallery feature\textemdash after which they could inquire about each of the probe. The study concluded with interviews gathering participants' perceptions and potential use cases for the application, confirming the formative study's findings, and exploring additional functionalities.

The data from the main study interviews were also translated and transcribed into English. All authors collectively reviewed these translations and made notes accordingly. These notes were analyzed along with the field notes to determine the emergent themes across the study's interaction component. We studied the data using reflexive thematic analysis proposed by Braun and Clarke \cite{clarke_successful_2013}. Table \ref{Table 5} presents the themes, sub-themes, and codes.

\subsection{Positionality}
Our research team, comprising Indian-origin researchers with insights into older adults' challenges, brings personal experiences that underscore the importance of supporting this demographic. Our background contributes to a deeper understanding of the nuances shaping older adults' technology adoption and intervention needs. All authors' lived experiences with older adults in the family, their societal interactions, and their experiences of living away, have also contributed to a deeper reasoning behind this work's motivation and the exploration of the sociocultural implications discussed in this work.

%% file: 4_findings.tex
\section{Findings}

Our findings are organized around three key themes. First, we captured the factors older adults in urban India consider critical to their aging experience by examining retirement transitions, health, and wellbeing. Second, we explored the role of technology in their lives\textemdash from surveys, interviews, and video probe responses on \textit{TIES}\textemdash which revealed their experiences, knowledge, and willingness to adopt TV as a wellbeing medium. Third, by assessing their current methods of social connectedness and support, we identified the challenges and barriers to technology use and the potential of \textit{TIES} in enhancing collective care models in India. We integrated the video probe findings into the specific themes they addressed to improve readability. These insights can guide the design of contextually aligned positive aging technologies for India's older population.

\subsection{Adapting to Life Post-Retirement: Challenges, Routines, and Wellbeing}

\subsubsection{Retirement Transitions}

In urban India, retirement is increasingly understood not as the termination of professional work but as a transformative phase devoted to pursuing personal fulfillment and a quieter lifestyle. Our study data reveal that many older adults embrace retirement as an opportunity to \textbf{reduce life’s responsibilities} and \textbf{engage in meaningful activities} such as learning new skills and exploring spiritual practices. Several participants expressed that the shift toward a more relaxed, peaceful existence was central to their post-retirement experience. For example, P20 (80, M) and P23 (70, M) emphasized adapting to a calm lifestyle, while P16 (60, M) and P14 (76, F) viewed retirement as a chance to pursue previously unattended interests.

\begin{quote}
    "After retirement we did not feel the need to go out. We visit villages sometimes. But nothing grand. That is, I want to have a peaceful life." - P20, (80, M)\\
    "I don't have the tension to wake up early or anxiety to be in office on time. Now, I am preparing to spend time with my family. I want to learn English, Sanskrit, and do agriculture." - P16 (60, M)\\
\end{quote}

However, this transition is not free of challenges. Issues such as \textbf{reduced physical mobility} and \textbf{diminished financial stability} emerged as significant concerns, adding complexity to the quest for autonomy. Several participants, such as P18 (81, F) and P21 (61, M), commented on how these challenges have reshaped their daily routines and personal aspirations.

\begin{quote}
    "I had to climb two sets of stairs. Here I am not able to do much now." - P18 (81, F)\\
    "Before retirement, life was more luxurious; now, there are worries in terms of financial status." - P21 (61, M)
\end{quote}

\subsubsection{Health and Wellbeing}

The interviews further expose a nuanced interplay between physical health and the perception of aging. Most older adults in our study reported a generally healthy lifestyle; yet for those aged 75 and above, \textbf{challenges related to mobility and independence} have become more prominent. A segment of the population appears to \textbf{accept the natural decline} in physical capabilities as an \textbf{inherent part of aging}\textemdash a perspective supported by several personal accounts. One participant (P19 (70, M)) recalled his struggle with a heart-related event, while another (P25 (67, F)) described a slowing pace of daily activities.

\begin{quote}
    "After some time I got block in my heart and then heart attack as well and operated with a stent. For 12 years since then, I've been fine." - P19 (70, M)\\
    "I was very brisk till 60. Now, my work is, 'uh', I've become a slow worker." - P25 (67, F)
\end{quote}

Beyond physical ailments, issues such as reduced mobility and \textbf{occasional memory lapses} (such as P27 (81, M)) have prompted many to adopt daily practices that bolster mental and emotional wellbeing. Several participants, such as P9 (78, F), have \textbf{woven spiritual routines} into their lives as a means of maintaining fulfillment and continuity despite the obstacles imposed by aging.

\begin{quote}
    "I had to move around for work. I am not able to walk as much now. So I've had to reduce it as much as possible. I forget to take [medicine] often." - P27 (81, M)\\
    "I do see TV, but mostly spiritual channels. I used to watch serials, but I've grown old now \textit{(gets tired easily)}. Sometimes I just turn it on and leave it on spiritual channel to listen to some slokas that's running. \textit{(even if not watching)}" - P9, (78, F)
\end{quote}

These reflections underscore that \textbf{regular functioning} and \textbf{personal independence} remain central to older adults’ overall wellbeing, while also signaling an opportunity for targeted interventions that support both physical and spiritual health.

\subsection{The Role of Technology and TV in Supporting Aging Adults}

\subsubsection{Technology Use and Resistance}
Our exploration into technology usage among aging adults in urban India reveals a wide spectrum of engagement. Participants with a background in education or a history of professional interaction with technology demonstrated greater openness, often incorporating \textbf{digital devices into their daily routines.} One participant explained her comfort with using technology, having learned from a family member during visits.

\begin{quote}
    "Yeah, yeah. I've got a printer, wifi printer and everything. Yeah. I'm comfortable (with using technological equipment), because, when I go there, I spend time (learning) there watching her. She (sister) taught me how to use it. Also we have to give the (visiting family) same environment. So when they are used to this sort of life, at least we have to provide something to them." - P8 (68, F)
\end{quote}

On the other hand, \textbf{difficulties adapting to new technologies} were strongly voiced by some participants. The expressed challenges reflect the broader issue of a \textbf{high cognitive load} associated with learning and integrating unfamiliar digital systems—an obstacle that could potentially \textbf{hinder independent adoption.} They felt like a burden when they had to impose on family members to learn new technologies.  P12 (83, F), P26 (76, F), and P27 (81, M) highlighted difficulties in adapting to new technologies and expressed \textbf{hesitance in seeking help} from family.

\begin{quote}
    "Because, after a certain age, learning things is very difficult. I'll say myself. Because there is no easy thing to adapt, have not taken that into account just yet." - P12 (83, F)\\
    "He \textit{(husband)} can call and ask \textit{(the daughter to learn)}, but we don't know if it will be difficult for them \textit{(burden to the daughter's family)}. Isn't it?" - P26 (76, F)\\
\end{quote}

Apart from these difficulties, it is common for older adults to\textbf{ fall prey to predatory practices} such as scams and fraudulent schemes. Past negative experiences among a few, such as P6 (74, M), made them stop using it altogether, considering the endeavor risky.

\begin{quote}
    "I was using it. But not anymore. Something happened. So this guy threatened me. He said, 'I have got your passwords, information, documents, and all sorts of things. If you don't give me three thousand bitcoins, I will ruin your life.'" - \textbf{P6 (74, M)}
\end{quote}

Collectively, these narratives call attention to the \textbf{pressing need for support systems} that address both \textbf{cognitive and practical challenges}. They advocate for a focus on technologies and learning mediums with which older adults are \textbf{already familiar}, thereby promoting a \textbf{smoother integration} of new technologies into their lives.

\subsubsection{Role of TV in Routine and Autonomy}

TV continues to occupy a central position in the daily lives of older adults, serving both as an entertainment hub and as a potential platform for innovative health and social interventions. Most participants saw potential in the TV-based intervention, recognizing its potential utility, particularly for families living apart. Participants' willingness to adopt such an intervention varied based on \textbf{personal needs and lifestyle preferences.} After viewing the video probes of various features, there was consistent excitement among all the participants to understand and sometimes even suggest improvements on the various features depicted through the probes. When asked about their willingness to adopt such features in their TV, approximately one-third of participants expressed interest in using selected features, with others remaining neutral but acknowledging possible benefits. For instance, those who are hard of hearing find sound notifications from the TV speaker helpful. They reasoned that such a feature \textbf{could improve the time spent conversing} as viewing someone on a larger screen could \textbf{retain the engagement} in a conversation.

\begin{quote}
    "Yes! I like it. If it is like that, it (video calling) will be more intimate than taking over the phone. Yes, when we have this, the conversation itself will go to another level. When we are on the phone, we just want to stop after a point (as it is hard to hear for me). We keep saying okay and cut the call. But this will probably make us have more conversation with people. I think." - P13 (85, M)
\end{quote}

Beyond enhancing communication, participants proposed that additional features\textemdash such as health tips and tailored spiritual content\textemdash could further personalize their TV experience. Such suggestions support a design that not only entertains but also functions as a supportive tool for daily health management. The idea of monitoring viewing patterns to reassure users that someone is checking on their well-being was popular.

\begin{quote}
    "This is good. Because if children are far away, they would want to know about parents health. With this it is possible. [proceeds to explain the story about the good morning message club to be the proxy for health]. Being able to check in a timely manner in terms of health is really good. That way this application is really nice." - \textbf{P25 (67, F)}
\end{quote}

Many appreciated the intervention's low cognitive load and familiarity, which were seen as key facilitators for adoption. The simplicity of the TV-based system was a crucial factor, with participants valuing its ease of operation and the minimal effort required to integrate it into their daily routines. Many older adults expressed comfort with \textbf{traditional smart TV remotes}, and the ability to use the intervention (with six buttons\textemdash Home, Menu, Back, Direction Dial, OK, Volume dial), underscoring the feasibility of a TV-based intervention to enhance autonomy without imposing additional cognitive demands.

\begin{quote}
    "I mean, I said it right. Generally, simple operation to senior citizens (---) it is a very nice thing. Otherwise too much complication is headache and trouble to the people. It should not make people get negative. That is why I like this." - P19 (70, M)
\end{quote}

These observations suggest that \textbf{familiar and intuitive interfaces} inherent in TV technology represent a promising avenue to support aging adults in maintaining their routine and personal autonomy.

\subsection{Community, Social Connectedness, and Support Requirements}

\subsubsection{Community and Social Connectedness}

The data indicate that established communication tools remain central to maintaining social bonds among older adults. While participants primarily used \textit{WhatsApp} for communication, often through video calls for distant family members, many still utilized platforms like \textit{Meet}, \textit{Zoom}, and \textit{Skype} as alternatives to connect with friends, family, and religious gatherings. Video calls were generally preferred, although some, like P5 (68, M), favored standard calls, citing issues with viewing angles or technical complexities of operation. Those with regular \textbf{in-person interactions} were \textbf{less inclined} to use new technologies, highlighting that family proximity influences technology choices.

\begin{quote}
    "He (spiritual guru) does only on Skype. Also, there is this thing called Meet right? My sons call me on weekends, we use that sometimes." - P12 (83, F)\\
    "Because of their children, they are forced to learn. For us (—) We aren't very keen on it." - P13 (85, M)
\end{quote}

Conversely, some participants lamented the \textbf{disconnect with newer digital platforms}, highlighting a \textbf{generational gap} in technology use. This gap restricts their capacity to fully leverage available social channels and \textbf{reinforces the appeal of familiar devices}\textemdash like the TV\textemdash as a medium for communication, which could \textbf{mitigate the cognitive burden} associated with learning new systems.

\begin{quote}
    "I've worked for so many years. But even with technology and all, I won't be able to keep in touch with those boys right. They all are from the technology age, they will say they are in facebook or instagram. But We don't have any such stuff. So don't keep in touch with them as much." - P13 (85, M)
\end{quote}

While messaging platforms like WhatsApp are commonly used, participants \textbf{expressed discomfort with their prolonged use.} The TV might better serve their communication needs by offering an interface they are already comfortable with and use regularly, thereby \textbf{integrating with their daily use} and \textbf{limiting the prolonged use} of other tools that cause them discomfort. These insights underscore the potential of TV-based communication tools, which can serve as a medium more in line with familiar practices, easing the cognitive demands associated with newer technologies.

\subsubsection{Facilitators and Barriers to Adoption}

The adoption of a TV-based intervention among older adults is underpinned by several notable facilitators. Many participants expressed strong appreciation for the \textbf{simplicity and intuitive nature} of the system, emphasizing that familiar elements\textemdash such as a large screen and a \textbf{traditional remote control}\textemdash integrate seamlessly into their daily routines. This ease of use minimizes the cognitive burden typically associated with new technologies and thus supports positive engagement. For instance, P9 (78, F), noted:

\begin{quote}
    "That's actually very nice. If it is bigger, can read it easily too. Very good. Of course, won't everyone be interested? (in using it) That sounds nice, I say. It will be helpful when I am sitting and watching TV. If it comes on the TV whenever I don't have my phone with me, it will be helpful." - P9 (78, F)
\end{quote}

Despite these robust facilitators, the data also reveal emerging barriers that, for some users, may gradually undermine the intervention's appeal. Barriers to adopting new technology included \textbf{physical mobility issues, cognitive load, and affordability}. One relatively mild concern is rooted in family support dynamics; certain participants feel that \textbf{proximity to family diminishes the immediate necessity} of such technology.

\begin{quote}
    "But if we get reminders when we watch TV, that is good. Because with more age, the more memory issues we get. But not that much (interest) for me. My son is nearby so there is no need to be worried, if anything, he comes anyway." - P22 (64, F)
\end{quote}

Moving to more significant issues, some older adults cautioned that they might only see the value in the intervention as their health declines further. These individuals acknowledged that while the system offers clear benefits, its relevance might increase only when cognitive or physical impairments become more pronounced.

\begin{quote}
    "If you ask me why, especially after an age, the memory related issues could build up. If there is some unusual activity, if they miss a routine, people will be afraid. Family caring for him should be able to get that information. 'uh' (pause) I don't think I need it just yet. But, only because I am still mobile. I have a routine and I don't leave much. If I start getting memory losses. Then I will definitely use it. The moment health starts declining, that's when I would need this." - P20 (80, M)
\end{quote}

Thus, while the inherent \textbf{simplicity, familiarity, and low cognitive load} of the TV-based system are \textbf{powerful facilitators} for its adoption, these positive factors must be balanced against barriers that may escalate as the user's physical and cognitive conditions change. The gradual emergence of these challenges calls for the intervention to be adaptable, ensuring \textit{continued relevance} as users \textit{transition through varying phases of aging}.

%% file: 5_discussion.tex
\section{Discussion}

As India embarks on a paradigm shift due to the increasing number of nuclear families, there is a growing need to support older adults aging in place. Our exploration of \textit{TIES}, a TV-based interactive design, unveils how familiar technology can serve as an effective conduit for enhancing wellbeing and quality of life for older adults. Against the backdrop of evolving societal expectations and infrastructural challenges in India, our findings underscore the potential of leveraging a trusted, accessible medium to foster social engagement, deliver health reminders, and support independent living. In this discussion, we articulate our theoretical and practical contributions, outline design recommendations for future interventions, and engage with ethical considerations in line with the broader CSCW discourse.

\subsection{Life Changes\textemdash Stark Differences of the Global South}
The lived experiences of older adults in India differ markedly from those documented in the global north, largely due to distinct socio-cultural and infrastructural dynamics. As evidenced in our findings, patriarchal norms and limited social infrastructure shape the post-retirement experience. In India, expectations for older adults to reside with the eldest son or within close familial networks continue to prevail, and extra-family care is stigmatized \cite{thampiAgingPlaceCommunityDwelling2024}, even as the trend toward nuclear families increases. This dissonance is associated with experiences of social stigma, as noted in recent news reports \cite{apIndiaGrowsOlder2024}, and reflects the significance of financial security, health, and self-determination as reported factors influencing wellbeing.”

Moreover, India’s collective care systems and constrained infrastructure, as discussed in studies \cite{ugargolFamilyCaregivingOlder2018a, ugargol_care_2016}, leave older adults more vulnerable to isolation and dependency. In this context, \textit{TIES}\textemdash though not a direct solution for physical limitations such as reduced mobility\textemdash can promote regular contact with family and provide actionable insights regarding overall wellbeing. In an environment marked by escalating safety concerns \cite{anvayaaSafetyConcernsGrowing2024} and multifaceted challenges \cite{kumarSeniorCitizensStatus, jaiswalIndiasElderlyCare2024}, our observations and findings indicate that a TV-based intervention with features like health monitoring, emergency support, and memory reminders may address several challenges related to aging in place within the Indian context.

Importantly, our findings also illuminate how familiar technology supports social connectedness. Participants who lived apart from family members expressed strong appreciation for TV-based care features that offer a non-intrusive means of staying connected. Conversely, those residing in close proximity to family showed less reliance on such interventions. This observation resonates with prior work on the integration of technology in daily routines \cite{schroederOlderAdultsNew2023, 10.1145/3392848} and the evolving caregiving roles in inter-generational households.

While Social Virtual Reality (SVR) has been explored in Western contexts \cite{10.1145/3359251}, its high cost and the limited tech-savviness of older adults in India hinder widespread adoption. Adoption challenges may also relate to cultural values around simplicity and detachment in later life, often informed by religious practices, as suggested in prior work \cite{tiwariIndianConceptsLifestyle2013}. Participants noted that post-retirement routines often led to social disconnection due to digital exclusion. This aligns with prior HCI research \cite{10.1145/3479524, 10.1145/3434166, 10.1145/3434166, 10.1145/3564855, 10.1145/3641027} on digital social interaction and older adults' reliance on familiar devices like TV to mitigate loneliness. The \textit{TIES} intervention addresses this by integrating video calling, messaging, reminders, and notifying into a familiar platform, reducing learning barriers.  

Some participants suggested its use during emergencies or lockdowns to enhance social ties within gated communities (drawing from their recent challenges, having just emerged from the COVID Lockdown), echoing findings from Xing et al. \cite{10.1145/3610032}. While this study focused on family care, the system's structure could theoretically support community engagement, an area that warrants further investigation. In the future, we could integrate community programs, virtual gatherings, or local activity alerts into the TV system with the intention of supporting connections beyond immediate family networks, contingent on future user engagement and design iterations. Drawing from CSCW literature, we recommend partnerships with many such local organizations (Not-for-Profit Organizations like Agewell India, HelpAge India, Abhoy Mission, Dadidada Foundation, and many more who have a history of working with others to further their cause) and healthcare providers to encourage active participation in both virtual and physical communities, aligning with CSCW’s focus on inclusive, community-driven technology.

\subsection{Technology as an Engagement Enabler: Familiarity over Bespoke}
Our exploration into technology use among older adults highlights that simplicity and familiarity are key enablers for sustained engagement. Participants emphasized that their daily interactions with technology are profoundly influenced by ease of use; they favor devices with intuitive interfaces and descriptive controls, such as smart microwaves or Amazon Firestick systems. This preference contrasts with a growing body of HCI research that advocates for specialized design and training for older adults \cite{10.1145/3170427.3170641, 10.1145/3491101.3503745, 10.1145/3290607.3299025, 10.1145/3411763.3441326, 10.1145/3313831.3376299, 10.1145/3411764.3445412, 10.1145/3411764.3445702}.

Underscoring that older adults are more receptive to interventions that build on the technology they already know and the existing infrastructure within their community, our data converge with Lambton-Howard et al.'s Unplatformed Design approach \cite{10.1145/3313831.3376179} and the Assets-based Design theory, based on Kretzmann's Assets-Based Community Development approach \cite{kretzmannAssetsBasedCommunityDevelopment1996}. Our observations regarding the necessity of understanding both the users' desires and needs, as well as the requirements of stakeholders beyond the direct users, align with the concept of adoption-centered design proposed by Chilana et al. \cite{chilanaUserCenteredAdoptionCenteredDesign2015}. Participants reported that using standard remote controls and familiar TV interfaces reduced perceived learning effort, suggesting a lower cognitive load compared to unfamiliar systems. Studies have indicated that such design strategies are associated with increased adoption and may support long-term usability among emerging user groups \cite{10.1145/3544548.3581000, 10.1145/3491102.3501949, 10.1145/3530190.3534824, 10.1145/3491102.3517575, 10.1145/3014362.3014367}, reinforcing the need to adapt existing systems rather than develop new ones for older adults.

Prior work \cite{alaoui_increasing_2012, dika_use_2015, davoodi_interactive_2021} explored TV-based solutions, though many required specialized infrastructure. Advances in smart TVs now enable cost-effective interventions via plugins like Roku TV, Google Chromecast, Amazon Firestick, and Apple TV. These reduce barriers related to complex interfaces \cite{10.1145/2702123.2702430}, UI/UX limitations \cite{10.3389/fpsyg.2017.01687}, and dexterity challenges \cite{10.1145/2702123.2702430}, making adoption easier through simple interactions. A sustainable, intuitive design approach has the potential to reduce learning effort, thereby facilitating continued use among older adults.

\subsection{Harnessing Video-Based Learning and Prototyping for Older Adult Technology Adoption}

The cognitive burden of learning new technologies and the accessibility of visual learning modalities emerged as a prominent theme across participant responses in our study, aligning closely with prior research on observational and collaborative learning in older populations \cite{doi:10.1080/036012799267846, 10.1145/3637378}. While Fan and Truong \cite{10.1145/3209882} outlined guidelines for older adult-friendly product instructions, their practical adoption remains limited. Our findings demonstrate that video-based usability testing provides older adults with intuitive, step-by-step exposure to novel features without necessitating immediate hands-on interaction, aligning with prior research on the benefits of visual explanations \cite{choi_2007, bobek_creating_2016}. Additionally, studies suggest that visual aids enhance technology adoption \cite{10.1145/2399193.2399195, 10.1145/3411764.3445702}, reducing cognitive overload and improving engagement.

Building on this visual learning preference, we introduced video probes as a non-intrusive tool for usability exploration. Unlike live demonstrations, video probes allow older adults to engage with technology concepts without the pressure of real-time interaction. Unlike the Wizard-of-Oz technique, this approach does not need to account for operator influence. Participants with physical or cognitive constraints appeared to benefit from this approach, which also offered a potentially scalable and cost-effective means to provide consistent experiences across sessions. Video prototypes provided a structured yet flexible way for participants to observe, reflect, and offer feedback on potential interventions, capturing nuanced preferences that might have been overlooked in traditional usability testing.

Moreover, video-based usability testing fostered equitable engagement in dyadic interviews. Since both participants watched identical scenarios, the process eliminated potential biases that arise when one person interacts with a prototype first, influencing the other's approach. Some participants, intrigued by interactive features, later engaged with high-fidelity \textit{Figma} prototypes, facilitating deeper discussions on functionality and feasibility. This method offered a balance between structured observation and open-ended exploration, which appeared to facilitate broader insights into technology adoption across diverse user groups.

Despite its advantages, video probes have limitations. While effective in eliciting preferences, they do not allow direct interaction, limiting participants' ability to assess usability in real-time. Future research could explore hybrid participatory methods\textemdash combining video probes with hands-on prototype testing\textemdash to capture both reflective and experiential dimensions of technology use. Such an approach aligns with CSCW’s emphasis on inclusive, iterative design, potentially enabling older adults to play a more active role in shaping technology that meets their evolving needs.

\subsection{Ethical Considerations and Technology Adoption: Balancing Usability with Privacy}

This study’s intervention integrates many features with TV catering to diverse needs, including vision and hearing impairments, indicating the usability of the intervention. These features were perceived by participants as enhancing safety and connectivity, and may offer pathways to improving access to healthcare, such as through linkages with local doctors and pharmacists. For those with sensory challenges, TV’s large, familiar interface adds significant value. Additionally, participants associated TV-based interventions with cognitive stimulation and improved mental wellbeing, aligning with prior work \cite{doi:10.1177/20552076231176648} suggesting such potential. Participants recommended prominent yet non-intrusive notifications, customizable to individual preferences, ensuring support without disrupting daily routines.

While participants responded positively to the intervention, ethical concerns emerged. Some worried about increased screen time reducing physical activity, though prior studies suggest this can be managed through Android-based features such as warning notifications/prompts, timer-based application lockouts, and nudges\cite{olsonNudgeBasedInterventionReduce2022}. Privacy concerns also arose regarding monitoring features, particularly who could access the data and how it would be used, reflecting similar concerns in Li et al.’s work on privacy vs. awareness \cite{10.1145/3610202}. Past research \cite{10.1145/3449212} has explored digital security for older adults, emphasizing the need for transparency and user control. Participants preferred an active role in managing their data, aligning with previous findings \cite{10.1145/3610029}. Adopting transparent advertising, commerce, and data governance models, as suggested by Aung et al. \cite{10.1145/3677113}, could help address these issues. For instance, opt-in consent mechanisms would allow users to decide who receives alerts, ensuring greater autonomy over their data. However, such mechanisms might increase the cognitive load involved with the operation and those need to be considered when the complete application is tested. Participants’ concerns suggest that increasing user autonomy and control may help mitigate these challenges. A balance must be struck between the choice of data storage, automated data clearance policies, and other mechanisms that do not require constant oversight while preserving user privacy and autonomous control. These considerations align with CSCW’s ongoing discourse on technology ethics and balancing usability with privacy \cite{10.1145/3476047, 10.1145/3610202, 10.1145/3610038, 10.1145/3637309}.  

While our design keeps health data local and user-controlled, future deployments must refine transparency, consent, and control mechanisms. As older adults become more familiar with TV-based systems, expectations around data privacy may evolve, potentially necessitating more sophisticated governance approaches. Periodic consent renewal prompts with minimal action requirements could help sustain continuous awareness of data-sharing preferences. This aligns with CSCW’s focus on ethical technology design, prioritizing user agency without compromising safety or usability. Additionally, to prevent over-reliance on technology, future studies could explore usage caps, mindfulness prompts, or activity reminders, encouraging a balance between digital engagement and physical interaction. By integrating these ethical safeguards, TV-based interventions may encourage more responsible use while preserving perceived benefits.

\subsection{Limitations and Future Directions}
While our study offers valuable insights into the adoption of familiar technology among older adults in urban India, several limitations merit consideration. Our sample primarily represents older adults from middle to upper-middle socioeconomic backgrounds with established access to digital services, limiting the generalizability of our findings to semi-urban or rural populations. However, given the increased availability of smart TVs at nominal prices, and the use of augments, like \textit{Firestick} and \textit{Roku}, we considered the findings to be applicable to a larger population. Another limitation pertains to potential selection bias, given that our participants were already familiar with TV-based systems. This could also be due to the fact that TVs are widely accepted and familiar devices that have been available to Indian families for generations. However, broader recruitment strategies that include individuals with limited exposure to digital technologies would provide a more comprehensive understanding of adoption barriers and inform the development of targeted training resources. Finally, our reliance on video prototypes constrained the ability of participants to interact dynamically with the system. Future work should incorporate hybrid methodologies that combine video probes with direct hands-on testing, thereby enriching the data on user experiences and aligning more closely with the participatory design principles advocated in CSCW.

Future research should consider partnerships with community centers, senior citizen clubs, and local NGOs to extend the reach of the \textit{TIES} system. Prior research \cite{abiddinNonGovernmentalOrganisationsNGOs2022, soniHarnessingPowerCollaboration2023, langmannLocalCommunityCapacity2024, linComeUsFirst2024} indicates the feasibility of such collaborations in enhancing community engagement. Exploring multilingual support for the TV system in the future would also improve inclusivity, particularly for regions with diverse linguistic needs or lower-income and low-literacy groups.

%% file: 6_conclusion.tex
\section{Conclusion}
Through \textbf{participatory design and video probes}, this study advances CSCW's efforts toward \textbf{inclusive technology design} by demonstrating how \textbf{leveraging familiar devices like TVs} can enhance the wellbeing and quality of life for older adults. By understanding the factors that older adults consider vital for their aging journey and exploring TV-based interventions to support those factors (such as social connectedness, health, and wellbeing), our findings underscore the importance of \textit{familiarity, simplicity, and autonomy} as essential principles for designing interventions tailored to aging populations. Our observations show that older adults \textbf{engage well with familiar and easily adaptable systems.} By addressing the practical, social, and ethical dimensions of TV-based interventions, this work highlights the potential of \textit{TIES} to fundamentally support the family-centric care responsibility model prevalent in the Global South. This research provides valuable insights for future projects focused on sustainable technologies to support older adults. Future work should focus on refining this approach and exploring the integration of community services within the TV-based intervention to further enhance social connectedness.